\documentclass[draft]{agujournal2019}
\usepackage{url} 
\usepackage{lineno}
\usepackage[inline]{trackchanges} 
\usepackage{soul}
\usepackage{amsmath}
\usepackage{changepage}
\usepackage{array}
\usepackage{booktabs}
\usepackage{caption}
\usepackage{xcolor}
\newcommand{\citet}[1]{\citeauthor{#1}~\citeyear{#1}}

\draftfalse

\begin{document}

%
%



\title{The role of gouge production in the seismic behavior of rough faults: A numerical study}
%
%




\authors{Miguel Castellano\affil{1}, Enrico Milanese\affil{2}, Camilla Cattania\affil{2}, and David S. Kammer\affil{1}}

 \affiliation{1}{Institute for Building Materials, ETH Zürich, Switzerland}
 \affiliation{2}{Department of Earth, Atmospheric, and Planetary Sciences, Massachusetts Institute of Technology, USA}





\correspondingauthor{=name=}{=email address=}


\setlength{\belowdisplayskip}{5pt}



\begin{keypoints}
\item We consider the effect of gouge production in earthquake cycle simulations on rough faults.
\item Gouge production causes heterogeneity in fracture energy, favoring partial ruptures and more variable recurrence intervals.
\item  The fault releases less seismic energy overall, favoring aseismic slip and creep-dominated nucleation.
\end{keypoints}

%
%

%
%


\begin{abstract}  

Fault zones mature through the accumulation of earthquakes and the wearing of contact asperities at multiple scales. This study examines how wear-induced gouge production affects the evolution of fault seismicity, focusing on earthquake nucleation, recurrence, and moment partitioning. Using 2D quasi-dynamic simulations integrating rate-and-state friction with Archard's law of wear, we model the space-time distribution of gouge and its effect on the critical slip distance. The study reveals a shift from single to multi-rupture nucleation, marked by increased foreshock activity. The recurrence interval undergoes two separate phases: an initial phase of steady increase followed by a secondary phase of unpredictable behavior. Finally, we observe a transition in the moment partitioning from faster to slower slip rates and a decrease in the moment released per cycle relative to the case where no gouge formation is simulated. This research sheds light on wear-driven mechanisms affecting fault slip behavior, offering valuable insights into how the evolution of gouge along a fault affects its seismic potential.

\end{abstract}

\section*{Plain Language Summary}
Our research explores the evolution of faults -- cracks in the Earth's crust where earthquakes originate -- as they undergo wear from friction. Through simplified computer models, we examine how gouge (the small rock particles created by wear) influences the initiation, frequency, and magnitude of earthquakes. Our findings indicate that as more gouge accumulates on the fault over time, the initiation process of earthquakes shifts, leading to an increase in precursory activity before a major earthquake. Additionally, the time between earthquakes changes, initially becoming longer before turning erratic. We also observed a gradual decrease in the total energy released by faults as they wear down, that is, faults with more gouge host earthquakes of relatively smaller magnitudes.

%
%

%


%
%
%
%

\section{Introduction}

Rough faults are known to produce large volumes of gouge during frictional slip \cite{powerRoughnessWearBrittle1988,wangWearProcessesFrictional1994,mcbeckDecryptingHealedFault2021,mollonLaboratoryEarthquakesSimulations2023}. 
This is due to their steep topography with asperities that interlock and produce gouge through wear
\cite{powerRoughnessWearBrittle1988,aghababaeiCriticalLengthScale2016,frerotMechanisticUnderstandingWear2018,milaneseRoleInterfacialAdhesion2020}. Sheared by tectonic stresses in the Earth's crust, natural faults often exhibit far-from-equilibrium behavior, withstanding large elastic energy gradients that they dissipate through the nucleation of frictional instabilities (i.e. earthquakes). As these instabilities unfold, some of the stored elastic energy is emitted in the form of seismic waves while most of it is dissipated through a set of different phenomena, such as frictional heating, crack propagation, inelastic deformation, off-fault damage \cite{nielsenFractureEnergyFriction2016,rechesEnergyfluxControlSteadystate2019,coccoFractureEnergyBreakdown2023} and other non-linear dissipation processes. These unstable fronts originate when the interlocking contact asperities are loaded beyond their stability threshold, triggering a positive feedback mechanism known as dynamic weakening. In this process, the more they slip, the weaker they become, leading to even larger slip in a self-reinforcing cycle \cite{sammisPositiveFeedbackMemory2002} that relaxes the fault by re-distributing the frictional stresses all along it.
    
The re-distribution of stress on the fault following instability is a well-known mechanism that has been modeled extensively using numerical methods \cite{lapustaNucleationEarlySeismic2003,aagaardConstrainingFaultConstitutive2008,radiguetSurvivalHeterogeneousStress2013, bruhatInfluenceFaultRoughness2020, cattaniaPrecursorySlowSlip2021, albertiniStochasticPropertiesStatic2021, scharNucleationFrictionalSliding2021}. However, gouge production by wear is also known to play an important role in the self-organization and maturation of natural faults \cite{duarteNoiseSlidingFriction2009,milaneseEmergenceSelfaffineSurfaces2019,ostapchukAcousticEmissionReveals2021, mollonLaboratoryEarthquakesSimulations2023}, yet most models of fault dynamics ignore its contribution. Both observational and experimental studies have pointed out the stabilizing effect of gouge on fault slip \cite{maroneFrictionalBehaviorConstitutive1990,maroneLaboratoryDerivedFrictionLaws1998,carpenterFrictionalPropertiesSliding2012, carpenterFrictionalHeterogeneitiesCarbonatebearing2014}, enabling velocity strengthening mechanisms that favor aseismic creep over dynamic weakening. Still, given the scale and complexity of the problem, most studies cannot afford to capture the long-term evolution of gouge accumulation and its effect on the overall seismic cycle. Although several numerical works have studied the effect of gouge on the frictional behavior of simulated faults \cite{moraNumericalSimulationEarthquake1998,mair3DNumericalSimulations2008, abeEffectsGougeFragment2009,ferdowsiSlideHoldSlideProtocolsFrictional2021, mollonLaboratoryEarthquakesSimulations2023}, few of them incorporate a mechanism for dynamic production of gouge coupled to frictional slip. Gouge is likely to play a key role in rupture dynamics by modulating the slip weakening distance $D_\mathrm{c}$ \cite{maroneScalingCriticalSlip1993}, which in turn affects the fracture energy and can determine rupture propagation and arrest.

\citet{bizzarriRecurrenceEarthquakesRole2010}
simulated dynamic gouge production in a spring-slider system, modelling its effect on $D_\mathrm{c}$, and found the recurrence interval to decrease with the amount of gouge, but they neglected the effect of spatial heterogeneity. Moreover, \citet{talModelingEarthquakeCycles2023} used geometric considerations to account for gouge creation through the removal of material at the interface and found that the magnitude of events would remain higher with wear than without it, but they did not model its effect on the critical slip distance. The effect of gouge on the critical slip distance is important because it shapes the fracture energy profile along the fault, with notorious consequences on fault-slip dynamics. Yet, a thorough analysis of the complex feedback mechanics that underlie such coupling between friction (through $D_\mathrm{c}$) and wear-induced heterogeneous gouge production is missing. 

Here, we use a well-established framework for simulating cycles of earthquakes \cite{segallRoleThermalPressurization2012,cattaniaPrecursorySlowSlip2021} to explore fault dynamics induced by frictional wear on rough faults and shed light on the role that frictional wear plays in the seismicity of a fault over cycles of earthquakes. In particular, we analyze the evolution of seismicity patterns on rough faults as a result of wear processes that account for the generation and accumulation of gouge material. With this approach, we investigate the effect of progressive gouge accumulation on the recurrence interval, the nucleation dynamics, and partition of the moment between seismic and aseismic components.  


\section{Numerical model}
\label{sec:numerical_model}
\subsection{Governing and constitutive equations}

\begin{figure}[ht]
    \centering
    \includegraphics[width=\textwidth]{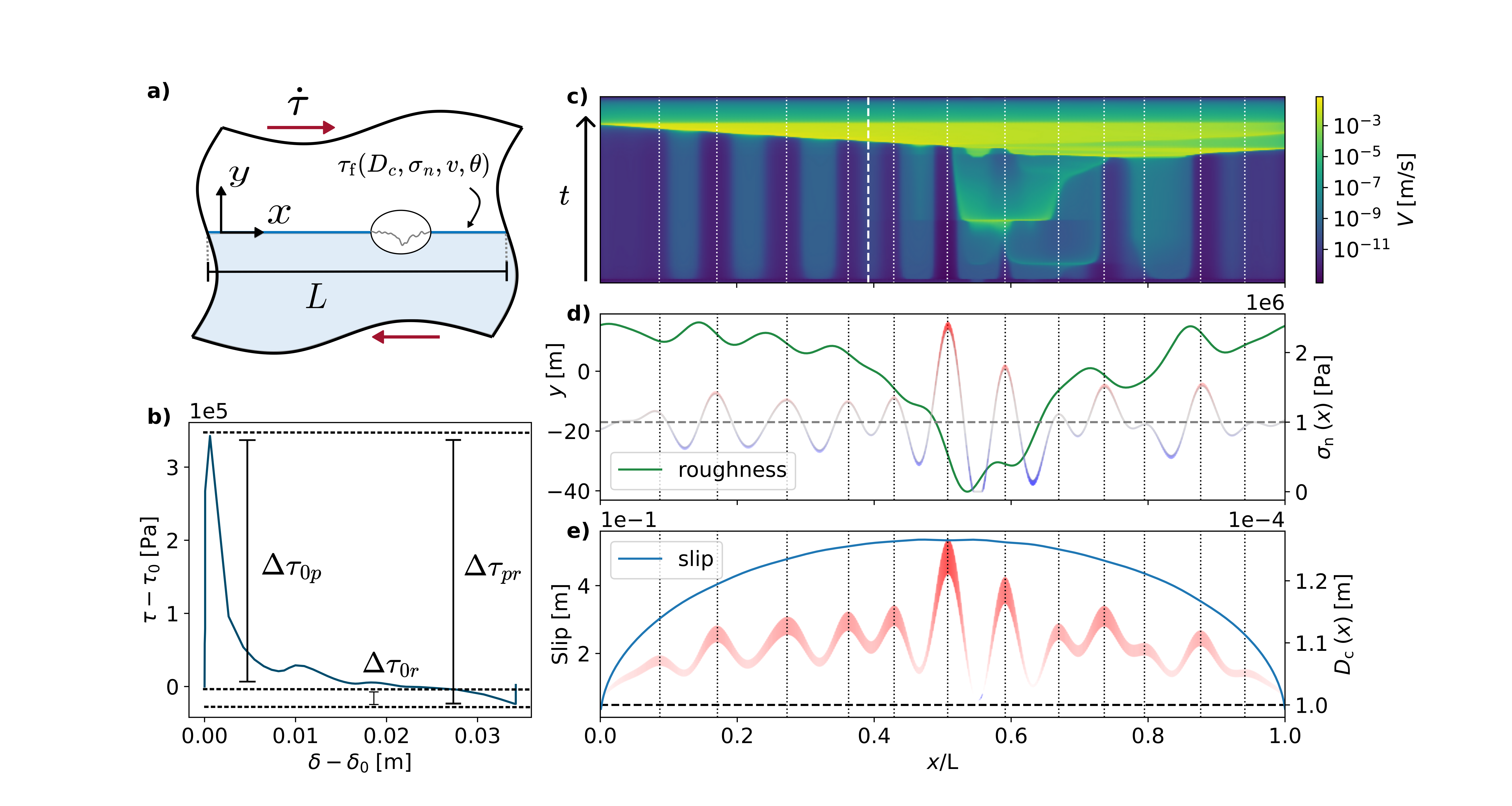}
    \caption{Simulation setup and sample data. \textbf{a}) Simulation setup. We impose a pure-shear fixed loading rate $\dot{\tau}$ on a rough fault of length $L$ with a rate ($V$) and state ($\theta$) constitutive law.   \textbf{b}) Frictional stress $\tau-\tau_0$ at a given point on the fault (marked in \textbf{c} with a white dashed line) as a function of the local slip $\delta-\delta_0$ for a given cycle, where $\tau_0$ and $\delta_0$ are the shear and slip at the beginning of the cycle ($t=t_0$). $\Delta \tau_\mathrm{0p}$ stands for initial-to-peak stress, $\Delta \tau_\mathrm{pr}$ peak-to-residual, and $\Delta \tau_\mathrm{0r}$ initial-to-residual stress changes. \textbf{c}) Space-time map of the slip rate during a given cycle. \textbf{d}) Example of one of the roughness profiles along with the associated normal stress distribution $\sigma_\mathrm{n}(x)$ and its evolution throughout the cycle (red shows increase, blue shows decrease). The horizontal dashed line marks the average background normal stress on the fault $\sigma_\mathrm{n}^0 = 1\mathrm{MPa}$. \textbf{e}) Slip profile at the end of the cycle (blue) and evolution of $D_\mathrm{c}(x)$ during the cycle as a result of the wear coupling (red). The shaded area corresponds to the difference between the start (brighter red) and the end (darker red) of the cycle. All panels refer to the same cycle.  Vertical dotted lines in panels \textbf{c}-\textbf{e} indicate stress peaks.}
    \label{fig:setup}
\end{figure}

We employ a quasi-dynamic boundary element code \cite{segallRoleThermalPressurization2012} to conduct 2-D plane-strain simulations of a rough fault (see Fig.\ref{fig:setup}a). Fault slip is governed by the following equation:
\begin{equation}
    \tau(x) - \tau_\mathrm{f}(x) = \frac{\mu}{2c_\mathrm{s}}V(x)
    \label{eq:quasi-dynamic}
\end{equation}
where $x$ is the distance along the fault, $\tau_\mathrm{f}$ is the frictional resistance, and $\tau$ is the shear stress due to remote loading and elastic transfer of stress between fault elements. The right-hand side is the radiation damping term, which represents the stress change due to radiation of plane S-waves, with $c_\mathrm{s} = 3700~\mathrm{m/s}$ the shear wave speed, $\mu = 30~\mathrm{GPa}$ the shear modulus, and $V(x)$ the slip rate.

Frictional resistance evolves according to rate-and-state friction \cite{dieterichModelingRockFriction1979,ruinaSlipInstabilityState1983} in its regularized form:
\begin{equation}
    \tau_\mathrm{f} (V,\theta) = a~\sigma_\mathrm{n} ~\text{asinh} \Bigl[  \frac{V}{2V_\mathrm{0}} \text{exp} \Bigl( \frac{\Psi}{a} \Bigr) \Bigr] 
    \label{eq:rate-and-state}
\end{equation}
where $\Psi = f_0 + b \ln(V_\mathrm{0}\theta/D_\mathrm{c})$, $\sigma_\mathrm{n}$ is the normal contact pressure, $f_0 = 0.6$ and $\mathrm{V}_0 = 10^{-6}~\mathrm{m/s}$ are reference values for the friction coefficient and the slip rate, $D_\mathrm{c}$ is the critical slip distance, $a = 0.015$ and $b = 0.020$ two constitutive parameters and $\theta$ is the state variable, which represents the local contact area of asperities, expressed in time-units as a proxy for the asperity `life-span'. For the state evolution, we employ the aging law \cite{dieterichModelingRockFriction1979, ruinaSlipInstabilityState1983}:
\begin{equation}
    \dot{\theta} = 1 - \frac{V\theta}{D_\mathrm{c}}.
    \label{eq:aging-law}
\end{equation}
This constitutive behavior causes stress to increase from the background stress $\tau_0$ with fault load reaching its peak strength at $\Delta \tau_\mathrm{0p}$ and then unload down to $\tau_0 - \Delta \tau_\mathrm{0r}$ over a slip distance proportional to $D_\mathrm{c}$ (see constitutive behavior of a fault-point in Fig.~\ref{fig:setup}b).

\subsection{Roughness}

To model fault roughness, we use the method from \citet{cattaniaPrecursorySlowSlip2021} by adopting a fractal surface geometry (see Fig. \ref{fig:setup}d), characterized by power spectral density
\begin{equation}
    P_\mathrm{h} = C_\mathrm{h} |k|^{-\beta}
\end{equation}
where we assume the shaping of the fault's geometry to be well described by a non-stationary process with stationary increments \cite{milaneseEmergenceSelfaffineSurfaces2019}. In such case, $\beta=2H+1$, where $H$ is the Hurst exponent. For natural faults, this is typically between 0.6 and 0.8 \cite{renardConstantDimensionalityFault2013,beelerScaleDependenceFaultSurface2023}; here, we set $H=0.7$. To keep the simulations feasible in terms of computation time, we only include wave numbers up to $k_{\mathrm{max}} = 2\pi(N_{\mathrm{max}}-1)/L$, where $L = 15.26$~km is the size of the fault (see Fig.~\ref{fig:setup}a) and $N_{\mathrm{max}}$ is set to $L/L_\mathrm{n} = 15$, $L_\mathrm{n} = 1.02$~km being the nucleation length, calculated from \citet{rubinEarthquakeNucleationAging2005} as
\begin{equation}
\label{eq:nucleation_length}
 L_\mathrm{n} = \frac{1}{\pi} \frac{\mu^{\prime} D_\mathrm{c}}{b\sigma_\mathrm{n}} \Bigl( \frac{b}{b-a}\Bigr)^2 ~,
\end{equation}
where $\mu^{\prime} = \mu / (1-\nu)$ and $\mathrm{\nu} = 0.25$ is the Poisson's ratio. Moreover, we define the minimum wavelength of the heterogeneous stress profile on the fault by $\lambda_\mathrm{min} = 2\pi/k_\mathrm{max} = 1/(N_\mathrm{max}-1)L$, where $N_\mathrm{max}$ is always fixed at 15. Therefore, $\lambda_\mathrm{min} = 1.09\; \mathrm{km}$, which is very close to the nucleation length $L_\mathrm{n} = 1.02 \; \mathrm{km}$ (Eq.~\ref{eq:nucleation_length}). This allows us to capture more explicitly the interaction between nucleating fronts and strength peaks (asperities) along the fault. \par

\subsection{Loading}

In order to simplify the loading process, we restrict the stress-rate tensor to a pure shear component, which is the one responsible for slip initiation. The resulting shear loading rate is given by
\begin{equation} 
\dot{\tau} = \frac{\dot{\sigma}_\mathrm{D}}{2}\sin ( 2\alpha_1 + 2\alpha_2),
\label{eq:shear-stressing-rate} 
\end{equation} 
where $\mathrm{\dot{\sigma}_D} = 0.25~\mathrm{MPa/year}$ is the differential stress rate, $\alpha_1$ is the loading angle, and $\alpha_2$ the slope of roughness. By setting the loading angle to $45^{\circ}$, we can fix the spatially averaged effective normal stress $\sigma_\mathrm{n}^0$, which we set to a uniform value of $1~\mathrm{MPa}$. The choice of a low value of $\sigma_\mathrm{n}^0$ is dictated by computational costs. Slip on a rough fault causes normal stress perturbations, and the time step is controlled by the smallest normal stress in the domain. Since negative stress changes could potentially induce fault opening if we assume a purely elastic response, we introduce a minimum value of $\sigma_\mathrm{min}$ below which the normal stress cannot fall, where $\sigma_\mathrm{min}=1~\mathrm{kPa} \ll\sigma_\mathrm{n}^0$. This threshold ultimately determines the simulation time step, introducing a limitation to the maximum amount of time that we can simulate in a reasonable run-time. In our case, the simulations model a period of around 30 years. By lowering the value of $\sigma_\mathrm{n}^0$ to $1~\mathrm{MPa}$ instead of $10~\mathrm{MPa}$, which would be more realistic, we are able to simulate around 35 to 40 earthquake cycles, while for $\sigma_\mathrm{n}^0 = 10~\mathrm{MPa}$, we only get around 4 in those same 30 years (see Supplementary Material, case HNS).

\subsection{Gouge production}

To model gouge production on the fault, we employ Archard's wear law \cite{archardContactRubbingFlat1953}. First developed in the context of adhesive wear for metal-on-metal contact, this linear model also captures well the formation of gouge in rock interfaces \cite{wangWearProcessesFrictional1994}, either by adhesive, abrasive or comminution mechanisms \cite{hughesFrictionalBehaviourWear2020}. Archard's model describes the total volume of wear produced as a linear function of the normal stress and the total amount of slip. To account for spatial heterogeneity, we apply it to every point in space ($x$) of our system so that the local wear rate in units of gouge thickness per unit time ($t$) is given by
\begin{equation}
    S_g(x,t) = K \frac{\sigma_\mathrm{n}(x) V(x,t)}{\eta},
    \label{eq:archard}
\end{equation}
where $K=10$ is a dimensionless wear coefficient of the same order of magnitude as the one in \citet{bonehGeotribologyFrictionWear2018}, $V(x,t)$ is the slip rate, and $\eta = 1.49~\mathrm{GPa}$ is the hardness, which is representative of soft materials like gypsum or calcite \cite{brozMicrohardnessToughnessModulus2006} that produce larger volumes of gouge as compared to harder materials like quartz. This allows us to shorten the time of the simulations, producing more gouge per sliding distance. In particular, we obtain an average gouge production rate of $6.7~\mathrm{mm}$ of gouge per meter of accumulated frictional slip, which can double for regions under higher normal stress, where $\sigma_\mathrm{n} > 1 \mathrm{MPa}$ (see Fig.~\ref{fig:setup}d). \par 
To quantify the effect of wear on the frictional response of the fault, we coupled it linearly to our rate-and-state friction model through the $D_\mathrm{c}$ parameter, in agreement with experimental observations \cite{maroneScalingCriticalSlip1993}. As a result, $D_\mathrm{c}$ evolves in space and time according to
\begin{equation}
    D_\mathrm{c}(x,t) = D_\mathrm{c}^0 + \mathrm{\gamma}  g(x,t),
    \label{eq:coupling}
\end{equation}
where $D_\mathrm{c}^0 = 10^{-4}~\mathrm{m}$ is the initial uniform value for $D_\mathrm{c}$, obtained from \citet{scholzCriticalSlipDistance1988} for deep faults, $\gamma = 0.01$ is a dimensionless coupling constant whose value is derived from the experimental work by \citet{maroneScalingCriticalSlip1993}, and $g(x,t)$ is the gouge field representing the thickness of the gouge layer. The latter 
evolves according to the classical transport or continuity equation 
\begin{equation}
    \dot{g}(x,t) = - \nabla(V(x,t) g(x,t)) + S_g(x,t)
\end{equation}
where the gradient term represents the outflow of gouge as a result of advection by the embedding velocity field, and $S_g$ is the source term, given in Eq.~\ref{eq:archard}. A representative evolution of $D_\mathrm{c}(x,t)$ during a cycle is characterized by a pronounced increase over regions with high normal pressure (see Eq.~\ref{eq:archard}) and accumulated slip (see elliptical slip profile in Fig.\ref{fig:setup}e). The advection of gouge is only responsible for a small percentage of the layer's thickness, but it does help to smooth out the distribution of $D_\mathrm{c}$, especially for higher normal pressures (See Fig. S8 in the Supplementary Material). \par

\section{Results and discussion}
\label{sec:results}

To isolate the effect of frictional wear and separate it from that of roughness, we ran two different sets of simulations, the reference simulations without wear contributions to the fault friction (i.e.~$\gamma=0$) and simulations implementing our proposed model (i.e.~$\gamma\neq 0$). Each set consists of 10 simulations, that differ only in the randomness of the fault's fractal geometry, introduced through random phase shifts in the frequency components of the power spectrum. That is, simulations within a given set all have different roughness profiles, but these profiles are the same across different sets to be able to compare them coherently and isolate the effect of wear. \par 

 Two extra sets of 10 simulations were run, where $\gamma = 0$ and  $D_\mathrm{c} = 3 \cdot \mathrm{D_c^0}$, for $\lambda_\mathrm{min}/L_\mathrm{n} = 1$ and $\lambda_\mathrm{min}/L_\mathrm{n} = 1/3$. This is done to study the differences between a situation where $D_\mathrm{c}$ is higher but uniform and another situation where $D_\mathrm{c}$ is increasingly heterogeneous due to wear. Furthermore, another 2 additional sets were run, with $\sigma_\mathrm{n}^0 = 10 \mathrm{MPa}$, to study the effect of normal stress on gouge formation (see Supplementary Material). In total, we ran 60 simulations, corresponding to six different cases (see Table 1).\par 

The resolution is maintained at $\Lambda_0/\Delta x > 2$ over the entire domain at all times, in agreement with \citet{cattaniaPrecursorySlowSlip2021}, where $\Lambda_0 \sim \mu D_\mathrm{c} / b\sigma_\mathrm{n}$ is the size of the quasi-static process zone at zero rupture speed \cite{andrewsRuptureVelocityPlane1976, rubinEarthquakeNucleationAging2005, perfettiniDynamicsVelocityStrengthening2008}. Finally, we define seismic events as slip events where the slip velocity anywhere on the fault exceeds the threshold velocity $V_\mathrm{dyn}$, here $\sim 0.4~\mathrm{cm/s}$, which depends on the normal stress on the fault as $V_\mathrm{dyn}=2a\sigma_\mathrm{n}^0 c_\mathrm{s}/\mu$ \cite{rubinEarthquakeNucleationAging2005}.\par 

\begin{table}[ht]
\centering
\begin{tabular}{
>{\raggedright\arraybackslash}p{2cm}| 
>{\centering\arraybackslash}p{1.5cm}
>{\centering\arraybackslash}p{1.5cm}
>{\centering\arraybackslash}p{1.5cm}
>{\centering\arraybackslash}p{1.5cm}
>{\centering\arraybackslash}p{1.5cm}
>{\centering\arraybackslash}p{2cm}}

\toprule
\textbf{Model} & \textbf{$\lambda_{\mathrm{min}}$} & \textbf{$\eta$} & \textbf{$\sigma_\mathrm{n}^0$} & \textbf{$L$} & \textbf{$L_{\mathrm{n}}$} & \textbf{$D_\mathrm{c}^0$} \\
\midrule
NG & 1.09 km & 0 & 1 MPa & 15.26 km & 1.02 km & $10^{-4}$ m/s\\
\midrule
G & 1.09 km & 0.01 & 1 MPa & 15.26 km & 1.02 km & $10^{-4}$ m/s\\
\midrule
NG-X1 & 3.27 km & 0 & 1 MPa & 45.79 km & 3.06 km & $3 \cdot 10^{-4}$ m/s\\
\midrule
NG-X2 & 1.09 km & 0 & 1 MPa & 15.26 km & 3.06 km & $ 3 \cdot 10^{-4}$ m/s\\
\midrule
NG-HNS & 109 m & 0 & 10 MPa & 1.52 km & 102 m & $10^{-4}$ m/s\\
\midrule
G-HNS & 109 m & 0.01 & 10 MPa & 1.52 km & 102 m & $10^{-4}$ m/s\\
\bottomrule
\end{tabular}
\caption{Different sets of simulations with corresponding parameters.}
\end{table}

\subsection{Nucleation dynamics: Single-rupture vs. Multiple-rupture}

\begin{figure}
\centering
    \includegraphics[width=\textwidth]{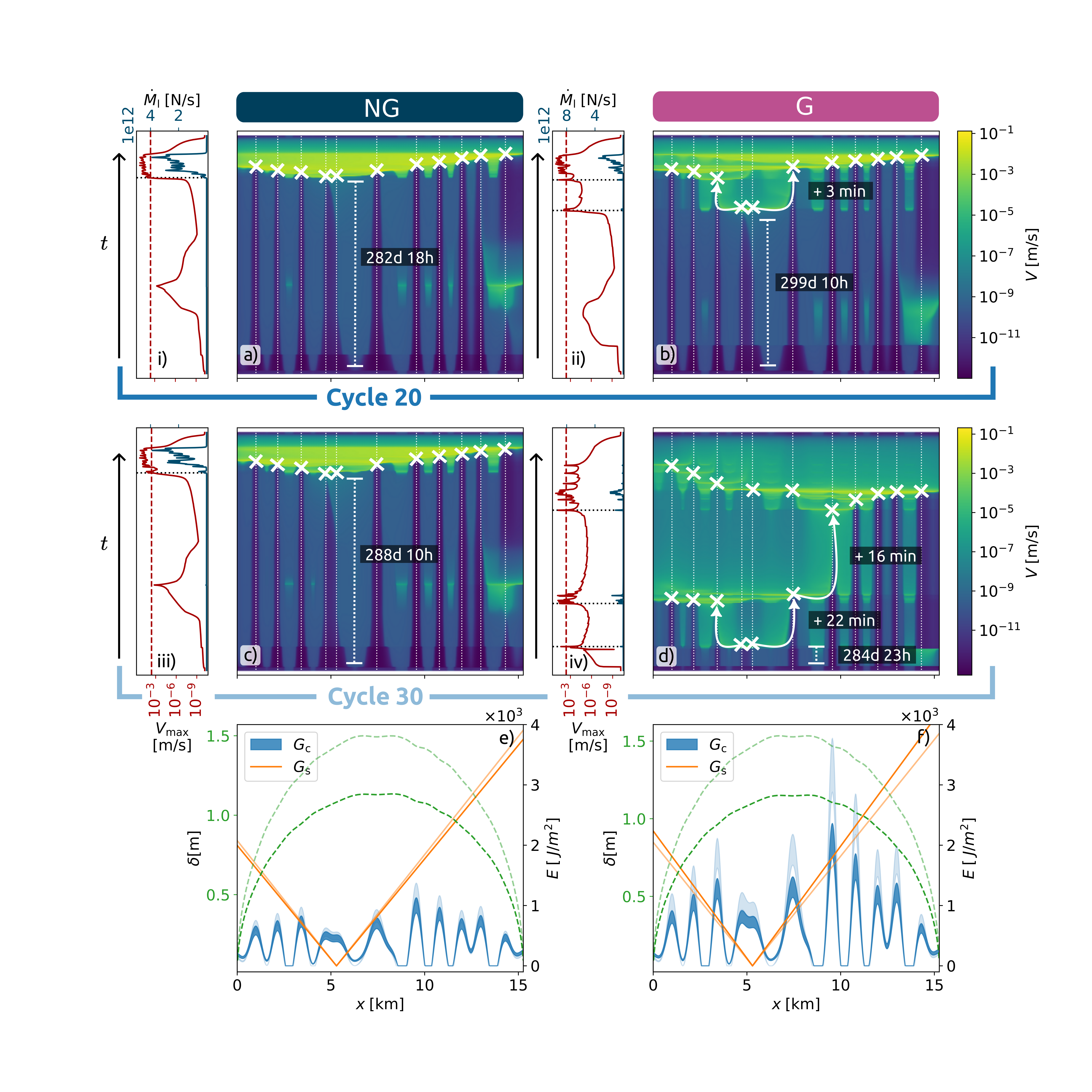}
    \caption{Earthquake nucleation process for representative cycles in simulations without (NG) and with (G) gouge production. Data for both cases is shown for an intermediate cycle (cycle 20) and a late cycle (cycle 30) of a given set of simulations.
    \textbf{i-iv}) Evolution of the maximum fault slip rate (red) and the moment rate per unit length (dark blue), where earthquakes are signaled with dotted lines as the peaks in the maximum slip rate going above the dynamic threshold and separated at least by one minute. \textbf{a-d}) Color maps of the slip rate distribution over the entire fault from the beginning to the end of the cycle. White 'x' signs mark the location of every asperity failure on the fault, which can trigger either a partial or a full rupture. \textbf{e,f}) Approximations of $G_\mathrm{s}(x)$ and $G_\mathrm{c}(x)$ (Eqs.~\ref{eq:Gc} and \ref{eq:Gs}) for both cases, along with the slip distribution $\delta(x)$ (dotted green lines). $G_\mathrm{c}(x)$ is plotted for a range of observed co-seismic slip rates going from $10^{-3}$~m/s to $10^{-4}$~m/s. Data from cycle 20 is shown in dark color, while that from cycle 30 is displayed in lighter color.}
    \label{fig:nucleation}
\end{figure}

We analyze the nucleation dynamics by considering the space-time distribution of the slip rate ($V(x,t)$) throughout different cycles, as illustrated in Figs.~\ref{fig:nucleation}a-d for an intermediate (Cycle 20) and a late cycle (Cycle 30) in both NG and G cases. \par

One of the main results of this analysis is that the presence of gouge (case G)  causes a progressive increase of slow slip events preceding the mainshock, together with a notable intensification of foreshocks; in contrast to the case without gouge (NG), where the earthquake initiation process features a system-spanning rupture. This does not imply that the NG case cannot exhibit a similar nucleation behavior to G. Indeed, \citet{cattaniaPrecursorySlowSlip2021} shows how, for sufficiently high normal stress perturbations, a fault can produce slow slip and foreshocks. However, through the positive feedback mechanism that links the stress and accumulated slip to the critical slip distance $D_c(x)$ by the formation of gouge, this process is enhanced. The situation is reversed in  \citet{talModelingEarthquakeCycles2023}, where they observe a reduction of aseismic slip when wear is accounted for. We believe this is due to the fact that they do not model the link between gouge formation and frictional traction at the interface, effectively ignoring the effect of wear on the fracture energy.  \par   

As the amplitude of both $D_\mathrm{c}(x)$ and $\sigma_\mathrm{n}(x)$ grows from cycle to cycle, enhanced by the accumulation of slip (see slip profile in Fig.~\ref{fig:setup}e), G goes from displaying only one small partial rupture followed by a full rupture in cycle 20 (Fig.~\ref{fig:nucleation}b) to a more complex process in cycle 30 (Fig.~\ref{fig:nucleation}d), involving two partial ruptures that break 60\% of the fault before triggering the full rupture. The magnitude of these partial ruptures can be deduced from the slip rate (red line in Figs.~\ref{fig:nucleation}i-iv) and the moment rate per unit length on the fault (blue line in Figs.~\ref{fig:nucleation}i-iv), which is computed as 
\begin{equation}
    \dot{M}_\mathrm{l} = \mu \int V(x,t)dx \quad .
    \label{eq:moment_rate}
\end{equation}
Prior to the mainshock, all simulations show an aseismic peak in the slip rate (Figs.~\ref{fig:nucleation}i-iii), while G-Cycle 30 shows two seismic peaks (Fig.~\ref{fig:nucleation}iv). As for the full rupture, the moment rate is much higher in cycle 20 than it is in cycle 30, indicating a decrease in magnitude (see blue lines in Figs.~\ref{fig:nucleation}ii and~iv). Both of these ruptures, partial and full, are prompted by the loading effect of expanding aseismic fronts that precede them. These fronts (indicated by white arrows in Figs.~\ref{fig:nucleation}b and d) seem to originate from small local events, which nucleate at the edge of weakening patches growing in between asperities. Very short-lived, these small events are self-arrested events, unable to penetrate the strength barriers ahead of them. Instead, they evolve into the aforementioned, stable and slowly growing aseismic fronts, before becoming unstable again as soon as they overcome such constraining barriers.  

The overall process of slip nucleation and propagation is described here in detail. As the fault is uniformly loaded, slip starts nucleating at the weakest parts, where the peak frictional strength ($\Delta \tau_\mathrm{0p}$) is lower. When they reach their peak strength, these weaker spots abruptly change their behavior, from strengthening to weakening as they start to slip, transferring all the excess stress outwards to their surroundings before reaching a plateau of strength (see Fig.~\ref{fig:setup}b) as they approach the residual level. The stress accumulating on the borders of the slipping area as a result of the redistribution of the external loading pushes the neighboring parts to start weakening too, creating a slip-weakening front around the patch that drives its own growth, while the central parts of the patch stay close to the residual strength level. After they reach a certain critical length (analogous to Eq.~\ref{eq:nucleation_length}), which depends on the local heterogeneities, these patches become unstable and start growing spontaneously, potentially propagating across the whole interface. This is what happens in NG (see Fig.~\ref{fig:nucleation}a, c), where the instability generated is strong enough to break through all asperities, triggering a propagating front that ruptures the entire fault. It is worth noting here that, as the instability unfolds, the parts that first started weakening immediately reach their weakening limit and go into approximately constant frictional strength. This becomes particularly important for the study of partial ruptures in the fault exhibited by case G (right column of Fig.~\ref{fig:nucleation}), which shows different nucleation dynamics that favor aseismic slip. A recent study \cite{castellanoNucleationFrictionalSlip2023} reported simulations in which frictional instabilities that arrested after hitting a barrier, led to a continuous stable expansion of the containing slipping patch with slow slip inside that loads the neighboring strong asperities until they break. This behavior has been reported in other numerical studies \cite{rubinEarthquakeNucleationAging2005,garagashNucleationArrestDynamic2012,dublanchetDynamicsEarthquakePrecursors2018, cattaniaPrecursorySlowSlip2021, saezFluiddrivenSlowSlip2024}, and it is what we observe too (see aseismic fronts indicated by white arrows on Figs.~\ref{fig:nucleation}b,d), suggesting it could be a leading mechanism for slow slip events and accompanying tremor (resulting from the self-arrested ruptures). \par 

To understand the nucleation mechanics at play, we can refer to the energy balance of nucleating patches, illustrated in Figs.~\ref{fig:nucleation}e-f. According to classical nucleation theory, a phase grows spontaneously whenever the change in free energy ($\Delta \mathcal{G}$) relative to its change in size becomes negative. Since $\Delta \mathcal{G} = G_\mathrm{c} - G_\mathrm{s}$,
where $G_\mathrm{c}$ is the fracture energy and $G_\mathrm{s}$ is the static energy release rate, the patch will spontaneously grow whenever $G_\mathrm{s} > G_\mathrm{c}$. These two terms are defined \cite{lawnFractureBrittleSolids1993,rubinEarthquakeNucleationAging2005} as
\begin{equation}
    G_\mathrm{c}(x) = \frac{\Delta \tau_\mathrm{pr}^2}{2W^*} = \frac{1}{2}b\sigma_\mathrm{n}(x)D_c(x)\Big[ \ln\Big(\frac{V_\mathrm{co}\theta_\mathrm{i}}{D_\mathrm{c}\Omega}\Big)\Big]^2
    \label{eq:Gc}
\end{equation}
\begin{equation}
    G_\mathrm{s}(x) = \frac{K^2}{\mu^{\prime}} = \frac{\pi \ell}{2\mu^{\prime}}\Delta \tau_\mathrm{0r}^2 =  \frac{\pi \ell}{2\mu^{\prime}}\Big[ \dot{\tau} T \Big]^2,
    \label{eq:Gs}
\end{equation}
where $W^*$ is the slip weakening rate, $K$ is the stress intensity factor, $\ell$ is the crack length, $T$ is the recurrence interval, $V_\mathrm{co}$ is the co-seismic slip rate, $\Omega = V\theta /D_\mathrm{c} \approx 1$ for steady-state and $\theta_\mathrm{i}$ is the state variable prior to the arrival of the crack-tip, which can be approximated by $T$ assuming $\dot{\theta}\approx 1$ interseismically. Note that the stress drop in eq.~\ref{eq:Gs} is not the coseismic stress drop during a single rupture, but the total shear stress imposed by loading since the previous full rupture. We assume that this is entirely released by the combination of the ongoing rupture and all previous earthquakes and creep within it, all of which contribute to the stress concentration outside the crack tip quantified by $K$.

As gouge builds up around the strongest points (high $\sigma_n$) surrounding nucleating patches, modulated by the amount of accumulated slip, and makes $D_\mathrm{c}$ larger (see Eq.\ref{eq:coupling}), the fracture energy $G_\mathrm{c}$ increases accordingly (see Fig.~\ref{fig:nucleation}f), making it harder for these fronts to propagate further. This increase in fracture energy is proportional to the total slip accumulated on the fault (dotted lines in Figs.~\ref{fig:nucleation}e-f), in agreement with Archard's law (Eq.~\ref{eq:archard}). As a result, the energy release rate of nucleating patches is not enough to overcome these barriers, as observed in Figs.~\ref{fig:nucleation}e-f, where arrest events tend to coincides with points on the fault where $G_\mathrm{c} > G_\mathrm{s}$, within uncertainties derived from the choice of co-seismic slip rate. For example, the rupture in the 20$^th$ cycle (NG) does not arrest, as predicted by the fracture mechanics criterion (Fig.~\ref{fig:nucleation}i,a,e); during the $30th$ cycle, a higher energy barrier causes a brief deceleration (Fig.~\ref{fig:nucleation}iii,e), which may then allow the rupture to overcome the barrier by lowering the fracture energy. In contrast, simulations with gouge are characterized by higher energy barriers which coincide with rupture arrest in the simulations (Fig.~\ref{fig:nucleation}b,d,f), with some minor discrepancies.

These arrested fronts then continue to slip aseismically ($\sim 10^{-6}\mathrm{m/s}$) within the arrest boundaries, as previously mentioned, expanding quasi-statically as the fault is loaded until they finally overcome the energy barriers, become unstable, and propagate seismically. This process would explain the transition over time towards more stable mechanisms for stress release, such as slow slip and micro-seismicity, akin to those observed in nature \cite{scognamiglioComplexFaultGeometry2018,tapeEarthquakeNucleationFault2018,mouchonSubdailySlowFault2023} and in the lab \cite{cebryCreepFrontsComplexity2022,selvaduraiModelingFrictionalPrecursory2023}. In contrast, the extra case NG-X1 features system-spanning events (as in NG) due to the spatial stretching of the fracture energy profile, while NG-X2 also exhibits a complex nucleation pattern with numerous partial ruptures like G (see Supplementary Material). This means that it is the shape of the fracture energy profile relative to the energy release rate that dictates nucleation dynamics on the fault. \par

To summarize, the introduced coupling between friction and wear opens up a new pathway for the fault's build-up of heterogeneity through $D_\mathrm{c}~(x)$, on top of the re-distribution of stress. This adds further complexity to the nucleation of events, which transition from a single full rupture to a complex mix of slow slip fronts and foreshocks, resulting in richer fault dynamics.

\subsection{Recurrence interval}

\begin{figure}
    \centering
    \includegraphics[width=\textwidth]{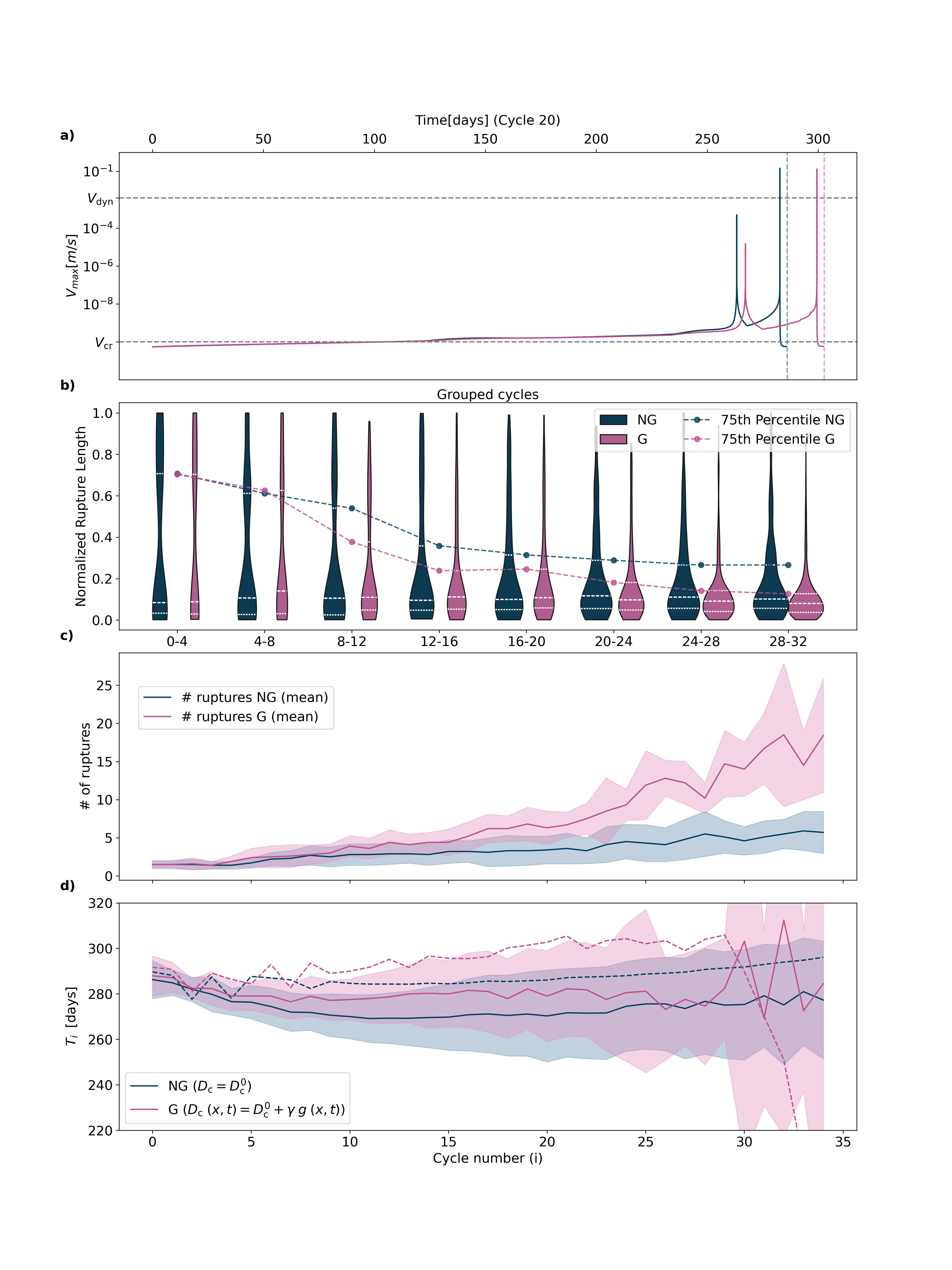}
    \caption{Evolution of the recurrence interval over multiple cycles. \textbf{a}) Maximum slip rate over time of intermediate cycle (20) for both cases. Vertical dashed lines show the end of the cycle, while horizontal lines indicate the creeping (bottom) and dynamic (top) threshold velocities. \textbf{b}) Evolution of rupture size distribution averaged over simulations in 4-sized groups, together with the $75^{\mathrm{th}}$ percentile. Rupture lengths are estimated as contiguous parts of the fault slipping above $V_\mathrm{dyn}$. \textbf{c}) Average number of ruptures per cycle. Shaded area shows the standard deviation. \textbf{d}) Evolution of the average recurrence interval across cycles for both NG and G. Dashed lines show specific cases corresponding to the set of simulations shown in \textbf{a}.}
    \label{fig:recurrence-interval-evolution}
\end{figure}

We examine the evolution of the recurrence interval ($T_i$) where $i$ is the cycle number. To measure it systematically, we identified the timestamps when the maximum velocity along the fault drops below the creeping velocity threshold $V_{\mathrm{cr}}$. This approach captures the entire cycle by detecting fault locking following a single rupture, or a sequence of closely spaced partial ruptures that collectively rupture the entire fault. Therefore, it spans the time from one post-seismic phase to the next, encompassing all intermediate phases of the seismic cycle, and is illustrated in Fig. \ref{fig:recurrence-interval-evolution}a for a given cycle in a simulation series. \par 


The time evolution of $T_i$ is seen to be affected considerably by the formation of gouge. In NG, it starts at 290 days, then drops and stabilizes at about 270 days after 10 cycles. Its behavior is quite predictable, with a notable increase in the standard deviation with time (see Fig.~\ref{fig:recurrence-interval-evolution}d). On the other hand, simulations with wear-friction coupling ($\gamma = 0.01$) show two distinct phases. Initially, $T$ diverges from the non-wear case (NG), stabilizing at a 10-day longer average. Subsequently (after around 25 cycles), it becomes shorter and more chaotic, with large variances observed in specific simulation sets. By the $30^\mathrm{th}$ cycle, it becomes quite erratic and unpredictable, showing pronounced fluctuations. This is in agreement with our previous reasoning about partial ruptures, which become more numerous with the number of cycles (see Fig. \ref{fig:recurrence-interval-evolution}b) and can alter the recurrence interval by changing the stress distribution of the fault mid-cycle. \par

The main explanation for these results comes from the growing heterogeneity in the fault as a result of the wear coupling. In spring-slider models that ignore the effect of spatial variability \cite{bizzarriRecurrenceEarthquakesRole2010, bizzarriWhatCanPhysical2012}, the recurrence interval is seen to decrease with gouge due to the effect of $D_\mathrm{c}$ on the critical stiffness of the spring ($k_\mathrm{cr}$). We observe a similar reduction of the recurrence interval in NG-X1 (see Supplementary Material). However, our results for G reveal the importance of heterogeneity. With the increasing amplitude of $D_\mathrm{c}~(x)$ and $\sigma_\mathrm{n}~(x)$ following the distribution of accumulated slip (Fig.~\ref{fig:nucleation}f) generated by the self-reinforced wear and stress transfer mechanisms fed by frictional instabilities, it becomes more difficult for slip fronts to propagate across the fault. This is because they have to overcome larger energy barriers. As a consequence, instead of a full rupture breaking the whole interface, partial ruptures become more frequent capturing part of the energy budget and saving it for later ruptures. We can clearly see this in Figs.~\ref{fig:recurrence-interval-evolution}b,c), where we count ruptures by taking the peaks of the maximum rupture length that coincide with peaks in the maximum slip rate going over the dynamic threshold. Hence, in the beginning, $T^\mathrm{G}$ becomes larger than $T^\mathrm{NG}$, since asperities are tougher, due to the accumulation of gouge and its effect on the fracture energy through $D_\mathrm{c}~(x)$ (see Eq.\ref{eq:coupling}), but the differences in fracture energy are not too large, so failure can propagate with ease. Basically, the failure mechanism of G is initially the same as that of NG but slightly delayed by the toughening of individual asperities. As fracture energy differences become larger, partial ruptures start to proliferate (Fig. \ref{fig:recurrence-interval-evolution}b,c), delaying the main shock of the current cycle \cite{turnerPartialRupturesCannot2024} but also creating stress concentrations where the main shock will eventually nucleate. \par 

When the energy released by the failure of the weakest asperities in the fault becomes too small compared to the fracture energy of the strongest and toughest ones, these ruptures fail to propagate throughout the entire interface (see Fig.~\ref{fig:nucleation}f). Instead, they are arrested by the tougher asperities, loading them for the next rupture to nucleate. In principle, one would expect that having multiple ruptures instead of a single full rupture would delay the end of the cycle, but in reality, it could also anticipate it, depending on the stress state of the fault. This would explain why the recurrence interval becomes so unpredictable. Initially, when the differences in fracture energy are not too large, $T$ depends exclusively on the fracture energy of the first asperity to fail, since this one unchains the instability of the whole interface (single full rupture). However, as the spread in fracture energy between asperities grows (see Fig.~\ref{fig:nucleation}f), the first failing asperity is not enough to break the entire interface anymore. This adds complexity to the recurrence interval, which starts to depend on the completion of multiple ruptures (see Fig.~\ref{fig:recurrence-interval-evolution}c), further contributing to the unpredictability. This behavior can also be observed in NG-X2 (see Table 1), where the recurrence interval is generally longer from the beginning, with a downward tendency, and then turns erratic after the $20^{th}$ cycle (see Fig.S5c in the Supplementary Material).  \par 

In conclusion, the evolution of the recurrence interval can be split into two phases: A growing phase (with respect to NG) and a chaotic phase. Initially, $T$ becomes longer than in the NG case as a result of individual asperities becoming tougher by increasing their critical slip distance, and full ruptures dominate. When the spread in fracture energy goes over a certain threshold, not all instabilities are able to break through the entire interface, making the recurrence interval dependent on multiple ruptures instead of one and, therefore, more unpredictable.

\subsection{Partitioning of the moment budget}

\begin{figure}
    \centering
    \includegraphics[width=\textwidth]{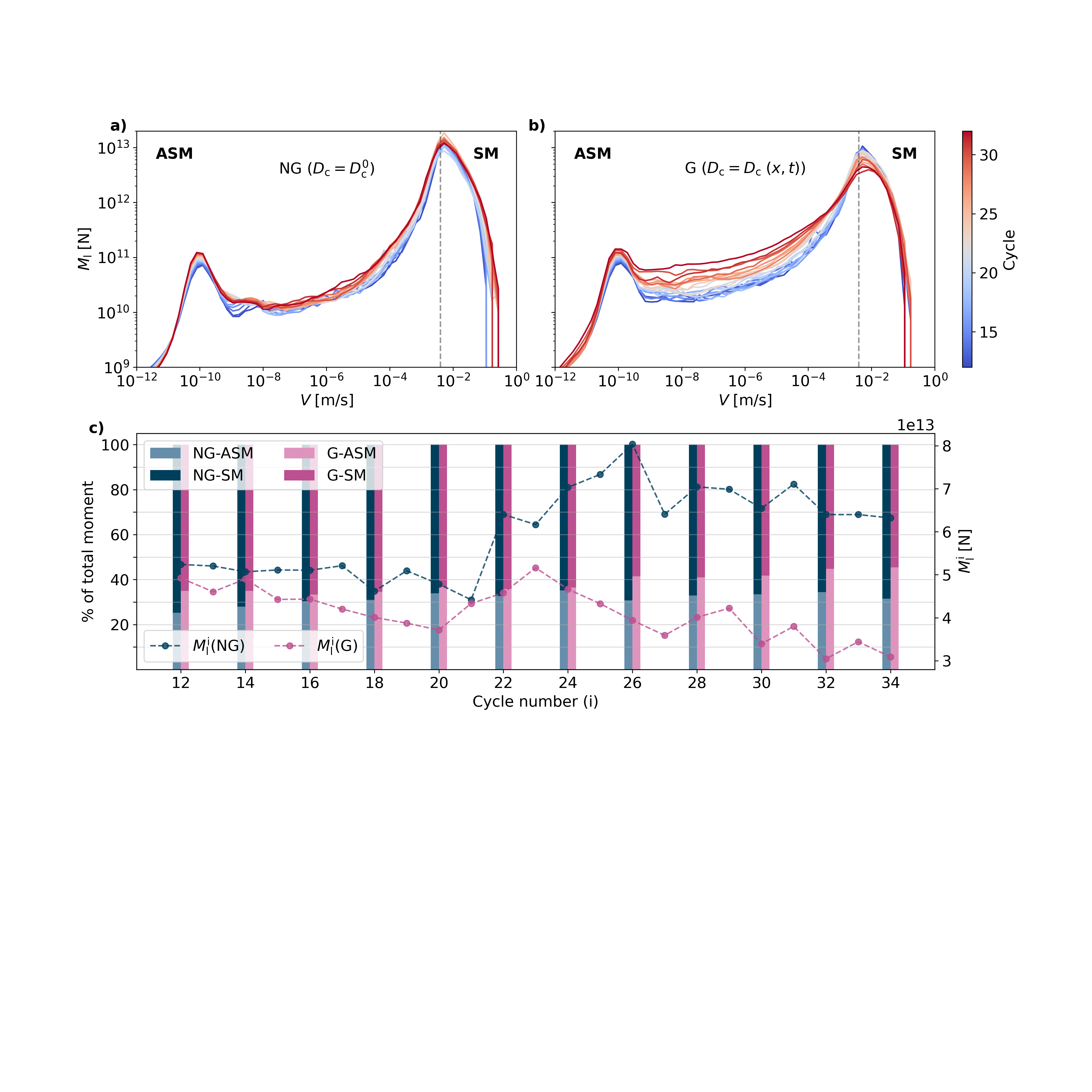}
    \caption{Comparative partitioning of the moment budget per unit length for both cases. \textbf{a, b}) Average distribution of the moment over a whole range of slip rates: aseismic (ASM) and seismic (SM), for cycles 12 to 34 and cases NG (a) and G (b). \textbf{c}) Evolution of both slip rate components (Aseismic [ASM] and Seismic [SM]) across cycles (left axis), as well as the moment per cycle per unit length (right axis), for both NG and G cases. 
    }
    \label{fig:moment-budget-partitioning}
\end{figure}

We also analyze the distribution of the moment budget between seismic and aseismic components and how it evolves with time. We compute the moment per unit length ($M_\mathrm{l}$) from the slip rate using
\begin{equation}
    M_\mathrm{l} = \mu \iint V(x,t) dx dt ~,
    \label{eq:M_l}
\end{equation}
and compute its distribution over the sliding velocity for both cases at different cycles averaged over the set of simulations (see Figs.~\ref{fig:moment-budget-partitioning}a,b).

Both NG and G show two peaks (or modes) that coincide with the creeping velocity $V_{\mathrm{cr}}$ and the dynamic velocity $V_\mathrm{dyn}$. The latter is used to split the moment in two ranges: aseismic (ASM) for $V<V_{\mathrm{dyn}}$, and seismic (SM) for $V \ge V_\mathrm{dyn}$. G shows, as expected, a clear shift in the slip-rate distribution of released moment from seismic to aseismic components between earlier (Fig.~\ref{fig:moment-budget-partitioning}a) and later cycles (Fig.~\ref{fig:moment-budget-partitioning}b). This is again explained by the rise of heterogeneity and complexity in the system, favoring more stable mechanisms for stress release, such as slow slip and micro-seismicity. \par 

Fig.~\ref{fig:moment-budget-partitioning}c shows, on its left axis, the partitioning of the fault's moment per unit length among aseismic (ASM) and seismic (SM) velocity ranges for 12 different cycles along the sequence. As expected, G experiences an increase of its aseismic component at the expense of the seismic one, which decreases by 10\% of the total moment. Conversely, the percentage of each component for NG remains rather constant, with the aseismic oscillating around 32~\%, as in NG-X1 (See Table 1), which has the same relative shape of $G_\mathrm{c}$ and $G_\mathrm{s}$. In NG-X2, however, the partitioning also remains constant but the aseismic component takes over with over 60\% of the total released moment. This is due to its squeezed fracture energy profile, which generates much more aseismic activity resulting from multiple rupture arrests. \par 

Finally, the right axis of Fig.~\ref{fig:moment-budget-partitioning}c shows the seismic moment per cycle and unit length ($M_l^i$), which is halved in G with respect to NG after the $24^\mathrm{th}$ cycle, corresponding to a 0.2 reduction in magnitude ($\Delta M_\mathrm{w} = -0.2$) over 12 cycles, according to Kanamori's formula for the moment magnitude scale \cite{kanamoriEnergyReleaseGreat1977}. This is in contrast to what is reported by \citet{talModelingEarthquakeCycles2023}, where they see that wear prevents the magnitude of events from going down. A possible explanation for this is that, contrary to how we model wear, where heterogeneity in the fracture energy always increases over time, their implementation has a homogenizing effect on the fracture energy, through the polishing of roughness and the redistribution of normal stresses at the interface.   


In summary, the enhancement of spatial variations in fracture energy as a result of gouge accumulation promotes rupture arrest on the fault, which in turn favors aseismic slip over seismic, reducing the amount of energy dissipated during the seismic cycle.

\section{Conclusion}
\label{sec:conclusion}

In this study, we explored how gouge production influences the mechanical behavior and seismogenic potential of rough faults. By integrating rate-and-state friction with Archard's model of wear, we examined the long-term impact on several key aspects of fault behavior. Our findings indicate that the recurrence interval of seismic events undergoes two distinct phases: an initial phase of steady increase followed by a secondary phase of unpredictability and chaos. We also observed a significant transition in the nucleation mechanics, shifting from a single full rupture mechanism to a more intricate, creep-dominated process involving increased microseismic activity. Additionally, this shift towards slower slip processes results in a notable reduction of the seismic moment per cycle released by the fault. This highlights the crucial role of frictional wear in altering fault dynamics over time.

\section{Open Research}

The simulation data generated in this study has been deposited in the ETH Research Collection database under accession code ethz-b-XXXXXX





\acknowledgments

This work was supported by an
ETH Zurich Doc.Mobility Fellowship. E.M. acknowledges funding from the Swiss National Science Foundation under grant P500PN 202863.



%
%



\bibliography{references}

%
%
%
%
%

\end{document}


%
%



\title{Supplementary material}
\title{The role of gouge production in the seismic behavior of rough faults: A numerical study}

\authors{Miguel Castellano\affil{1}, Enrico Milanese\affil{2}, Camilla Cattania\affil{2}, and David S. Kammer\affil{1}}

 \affiliation{1}{Institute for Building Materials, ETH Zürich, Switzerland}
 \affiliation{2}{Department of Earth, Atmospheric, and Planetary Sciences, Massachusetts Institute of Technology, USA}

\correspondingauthor{=name=}{=email address=}

\renewcommand{\thefigure}{S\arabic{figure}}
\setlength{\belowdisplayskip}{5pt}

\section{Description}

In this addition to the main article, we show four extra cases, two with no gouge formation (NG-X1 and NG-X2) where we set the critical slip distance to a higher value $D_c = 3D_c^0$, keeping it uniform, and another two where we increase $\sigma_\mathrm{n}^0$ to 10~MPa, with (G-HNS) and without (NG-HNS) gouge formation (See Table 1 in main text). We do this to compare with the case where $D_c$ grows heterogeneously (G case) and to study the effect of normal stress on gouge formation. In NG-X1, we make the fault three times longer than NG to keep $L/L_\mathrm{n}$ constant, since $L_\mathrm{n}$ depends linearly on $D_\mathrm{c}$ (see Eq.~5 from the main manuscript), while in NG-X2, we keep the same length of the fault that we used in NG.

\section{NG-X1}

\begin{figure}
\centering
    \includegraphics[width=\textwidth]{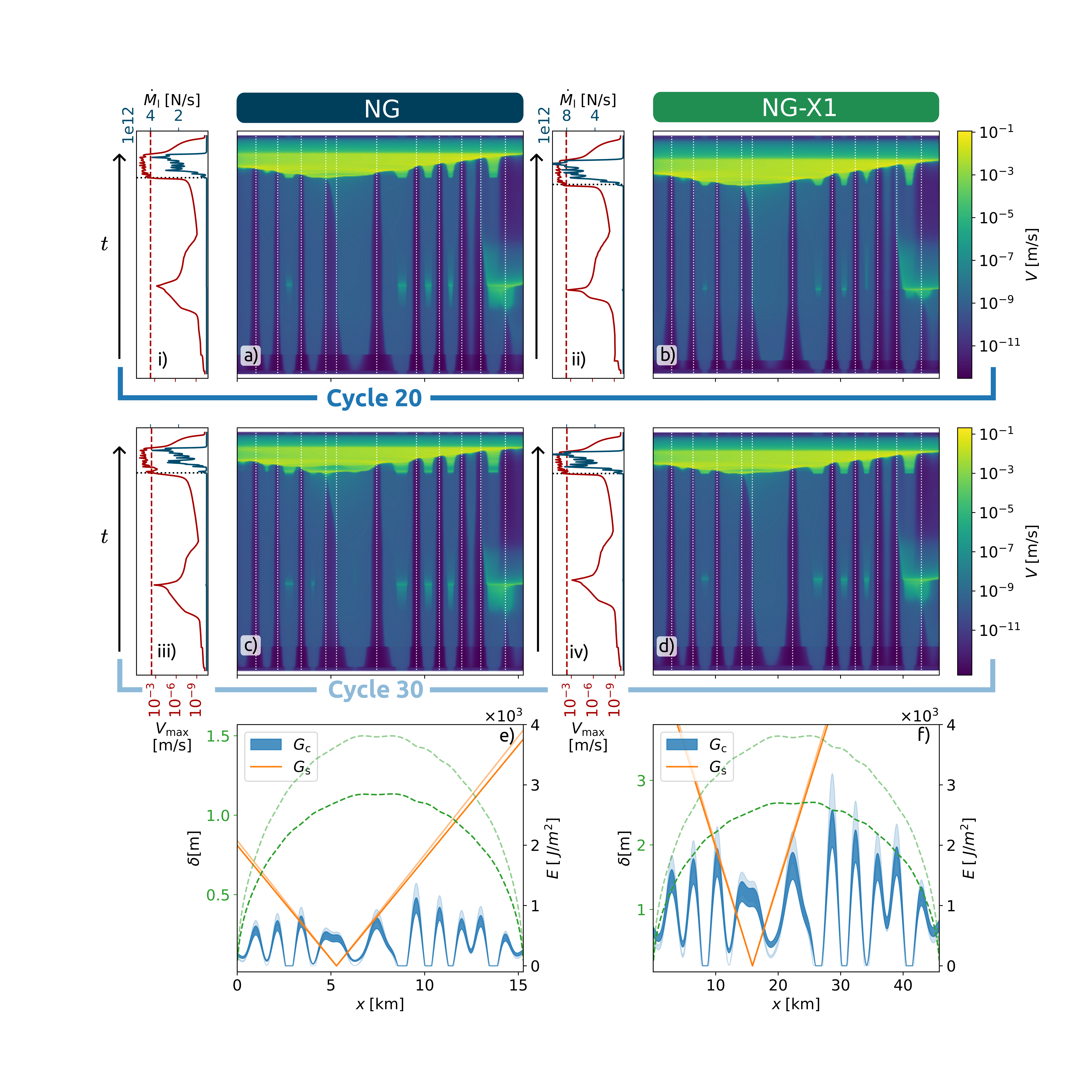}
    \caption{Earthquake nucleation process for representative cycles in NG and NG-X1. Data for both cases is shown for an intermediate cycle 20 (a,b) and a late cycle 30 (c,d) of a given set of simulations.
    For each cycle, the evolution of the maximum fault slip rate (red) and the moment rate per unit length (dark blue) are shown in (i-iv), where earthquakes are signaled with dotted lines as the peaks in the maximum slip rate going above the dynamic threshold and separated at least by one minute. The color maps (a-d) show the evolution of the slip rate distribution over the entire fault from the beginning to the end of the cycle. At the bottom (e,f), approximations for $G_\mathrm{s}(x)$ and $G_\mathrm{c}(x)$ are plotted for both cases, along with the slip distribution $\delta(x)$ (dotted green lines). $G_\mathrm{c}(x)$ is plotted for a range of observed co-seismic slip rates going from $10^{-3}$~m/s to $10^{-4}$~m/s. Data from cycle 20 is shown in dark color, while that from cycle 30 is displayed in lighter color.}
    \label{fig:nucleation-X1}
\end{figure}

\begin{figure}
    \centering
    \includegraphics[width=\textwidth]{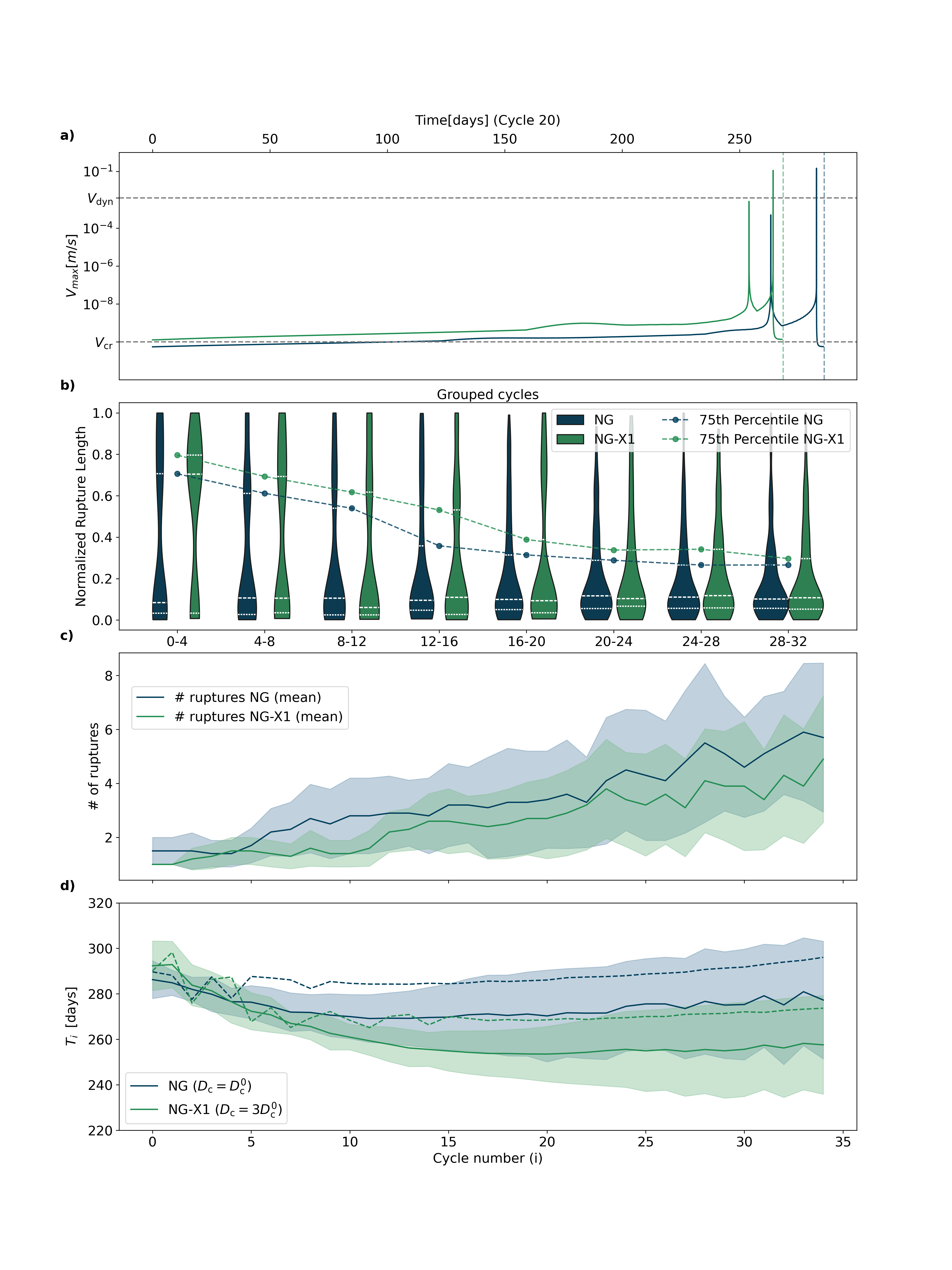}
    \caption{Evolution of the recurrence interval over multiple cycles. \textbf{a}) Maximum slip rate over time of intermediate cycle (20) for both cases. Vertical dashed lines show the end of the cycle, while horizontal lines indicate the creeping (bottom) and dynamic (top) threshold velocities. \textbf{b}) Evolution of rupture size distribution averaged over simulations in 4-sized groups, together with the $75^{\mathrm{th}}$ percentile. \textbf{c}) Average number of ruptures per cycle. Shaded area shows the standard deviation. \textbf{d}) Evolution of the average recurrence interval across cycles for both NG and NG-X1. Dashed lines show specific cases corresponding to the set of simulations shown in \textbf{a}.}
    \label{fig:recurrence-interval-evolution-X1}
\end{figure}

\begin{figure}
    \centering
    \includegraphics[width=\textwidth]{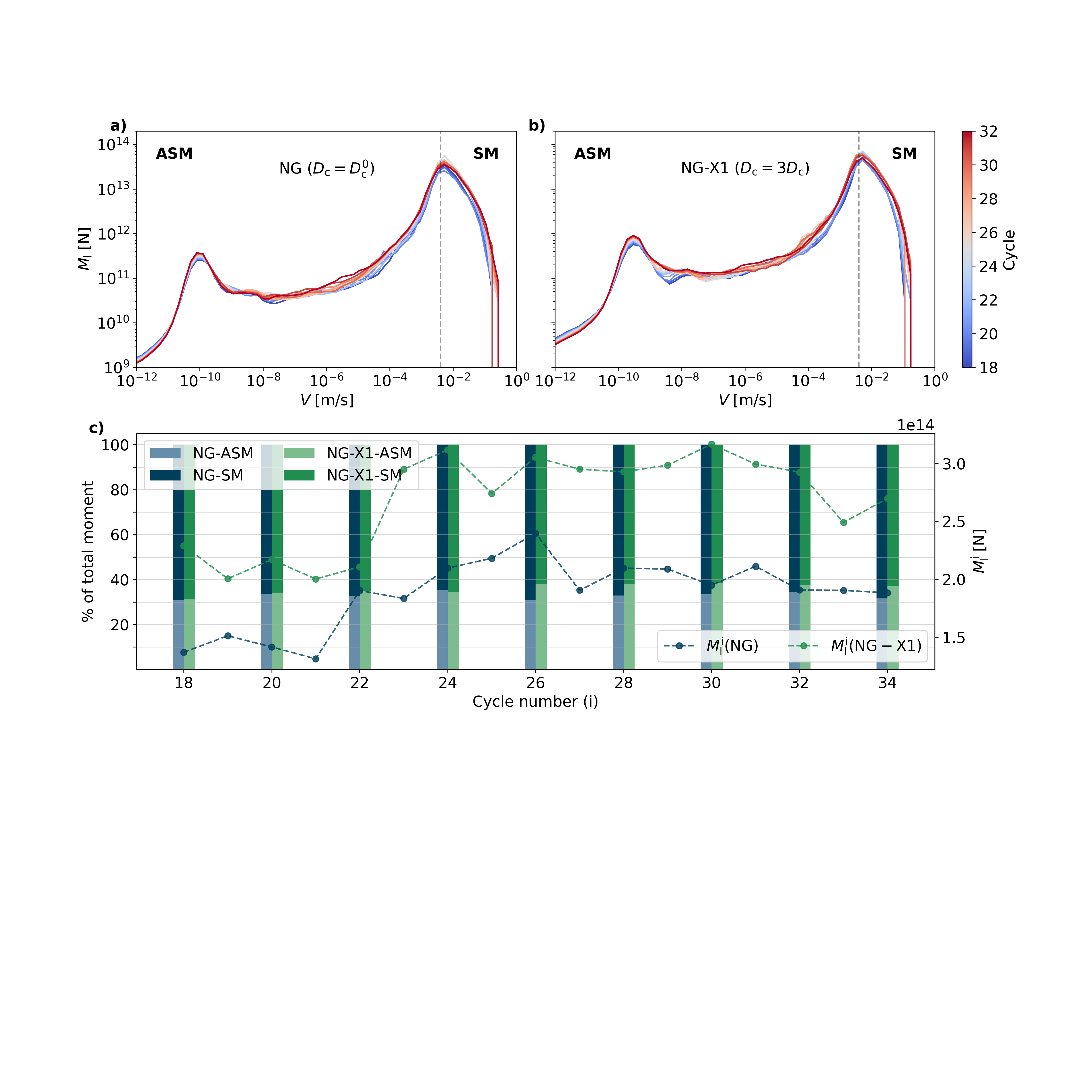}
    \caption{Comparative partitioning of the moment budget per unit length for both cases. \textbf{a, b}) Average distribution of the moment over a whole range of slip rates, aseismic (ASM) and seismic (SM) for cycles 12 to 34 and cases NG (a) and NG-X1 (b). \textbf{c}) Evolution of both slip rate components (Aseismic [ASM] and Seismic [SM]) across cycles (left axis), as well as the moment per cycle (right axis), for both NG and NG-X1 cases. 
    }
    \label{fig:moment-budget-partitioning-X1}
\end{figure}

In this case, the observed behavior is similar to NG. Since the nucleation length $L_\mathrm{n}$ is tripled by the effect of $D_\mathrm{c}$, we adjust the fault length accordingly so that the nucleation length $L_\mathrm{n}$ remains at $1/15$ of the total length $L$, while keeping the same resolution in the discretization as we did for NG. As a result, the minimum wavelength of the roughness profile, defined as $\lambda_{min} = 1/(N_\mathrm{max}-1)L$ where $N_\mathrm{max}=15$, is $3.27\; \mathrm{km}$, while only $1.09 \;\mathrm{km}$ for NG. The fracture energy is scaled by three, but the wavelength over which it does so also increases by the same factor, which keeps the relative shape of $G_\mathrm{s}$ and $G_\mathrm{c}$ constant (see Fig.~\ref{fig:nucleation-X1}e,f). Therefore, $G_\mathrm{s}$ stays always above $G_\mathrm{c}$, and the observed behavior resembles quite closely the one displayed by case NG (See Figs.~\ref{fig:nucleation-X1},\ref{fig:recurrence-interval-evolution-X1},\ref{fig:moment-budget-partitioning-X1}). In addition, the differences in fracture energy between asperities are not too large, further favoring unstable slip fronts to propagate easily throughout the entire system in single full ruptures (see Fig.~\ref{fig:nucleation-X1}). \par

When it comes to the recurrence interval (See Fig.~\ref{fig:recurrence-interval-evolution-X1}), this case shows a reduction over time, starting slightly above NG and stabilizing at a recurrence interval that is approximately 20 days shorter than NG after 15 cycles (See Fig.~\ref{fig:recurrence-interval-evolution-X1}d). In addition, the number of ruptures per cycle evolves similarly for both cases (see Fig.~\ref{fig:recurrence-interval-evolution-X1}c), as does their size distribution (see Fig.~\ref{fig:recurrence-interval-evolution-X1}b). \par 

Finally, the partitioning of the moment does not show any clear trend (Fig.~\ref{fig:moment-budget-partitioning-X1}). The ratio between seismic and aseismic components seems to stay rather constant throughout the sequence, with aseismic stress release accounting for around 32 \% of the total moment for both cases (Fig.~\ref{fig:moment-budget-partitioning-X1}c). Still, the moment released per cycle by NG-X1 is 50\% larger than that released by NG, on average, which can be partly attributed to the larger fault length, since moment is released over a larger area, but also to the higher creeping velocity ($V_\mathrm{cr}^\mathrm{NG-X1} = 3 V_\mathrm{cr}^\mathrm{NG}$) (see shifted left peak in Figs.\ref{fig:moment-budget-partitioning-X1}a, b), which contributes to a higher moment released per unit length by creeping patches (see Eq.~13 in main text). 

\section{NG-X2}

\begin{figure}
\centering
    \includegraphics[width=\textwidth]{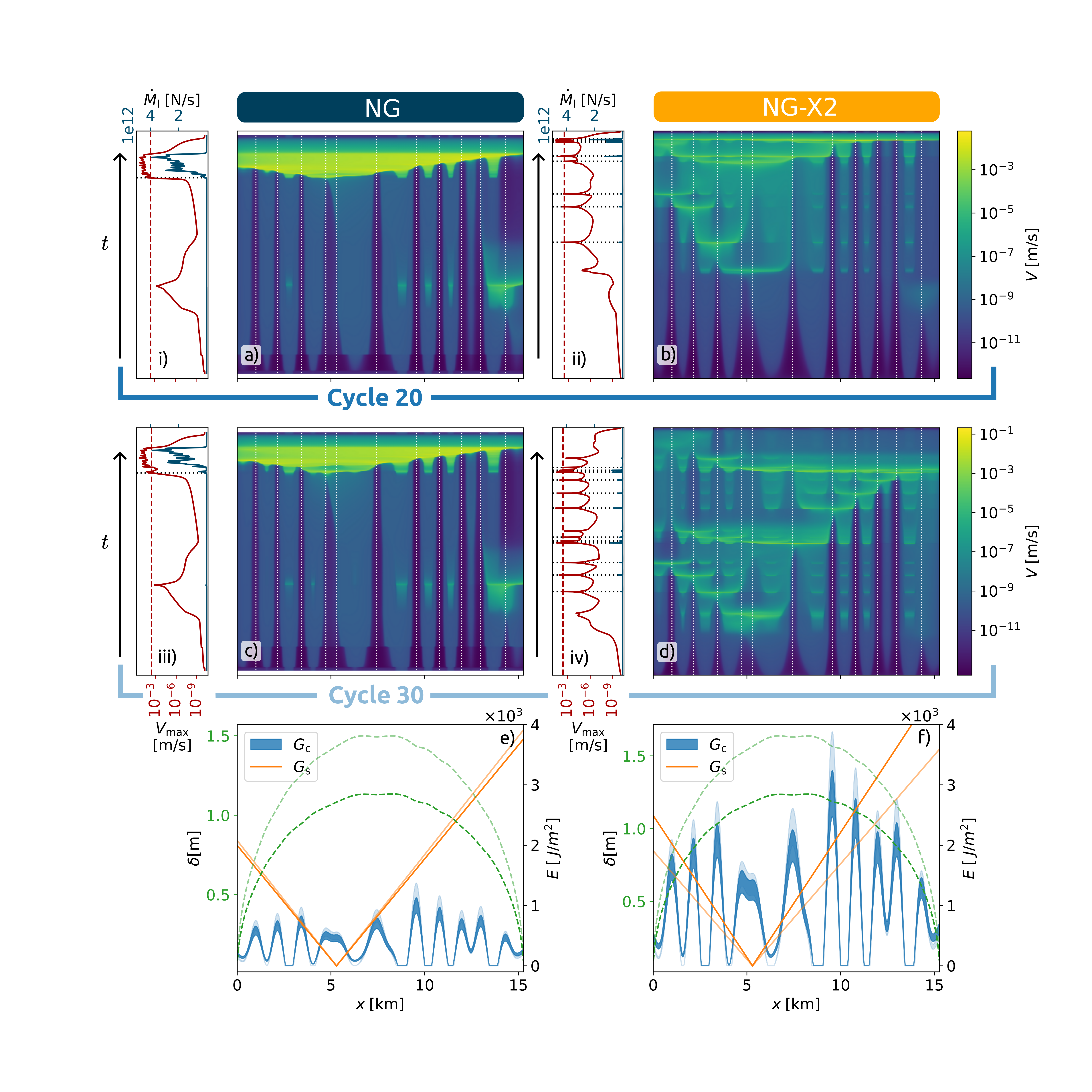}
    \caption{Earthquake nucleation process for representative cycles in NG and NG-X2. Data for both cases is shown for an intermediate cycle 20 (a,b) and a late cycle 30 (c,d) of a given set of simulations.
    For each cycle, the evolution of the maximum fault slip rate (red) and the moment rate per unit length (dark blue) are shown in (i-iv), where earthquakes are signaled with dotted lines as the peaks in the maximum slip rate going above the dynamic threshold and separated at least by one minute. The color maps (a-d) show the evolution of the slip rate distribution over the entire fault from the beginning to the end of the cycle. At the bottom (e,f), approximations for $G_\mathrm{s}(x)$ and $G_\mathrm{c}(x)$ are plotted for both cases, along with the slip distribution $\delta(x)$ (dotted green lines). $G_\mathrm{c}(x)$ is plotted for a range of observed co-seismic slip rates going from $10^{-3}$~m/s to $10^{-4}$~m/s. Data from cycle 20 is shown in dark color, while that from cycle 30 is displayed in lighter color.}
    \label{fig:nucleation-X2}
\end{figure}

In this case, we set $D_\mathrm{c} = 3 D_\mathrm{c}^0$, leaving the fault length as in NG and equal to 5 times the nucleation length $L_\mathrm{n}$. This means that we triple the fracture energy ($G_\mathrm{c}$) barriers through $D_\mathrm{c}$ as in NG-X1, but the wavelength ($\lambda_\mathrm{min}$) is only 1/3 of what it was in NG-X1, so it is shrunk in the x direction, making the crossings with the energy release rate ($G_\mathrm{s}$) more frequent (see Fig.~\ref{fig:nucleation-X2}f). As a result, the energy released by unstable fronts propagating through the fault is not enough to overcome the fracture energy barriers encountered, creating a nucleation pattern of multiple partial ruptures, which we can observe in Figs.~\ref{fig:nucleation-X2}b, d. However, the main difference with the G case, from the main text, lies both in the initial magnitude of the fracture energy barriers, which are much bigger, and in the evolution of their spatial differences with time. In G, these grow through the local increase of $D_\mathrm{c}$ with gouge, which adds on top of the increase in amplitude of normal stress perturbations due to re-distribution, while in this case (NG-X2), only the normal stress perturbations contribute, which can only account for so much growth, even though they are amplified by a larger $D_\mathrm{c}$. Therefore, the spatial evolution of the fracture energy is not so accentuated in this case, although the overall dynamics resembles a more advanced case G (after many more cycles). 

In terms of the recurrence interval, we observe a much higher value than NG during the first cycles (see Fig.~\ref{fig:recurrence-interval-evolution-X2}d), which is expected, since the fracture energy to break the first asperity is much larger, and nucleation is still done through a single rupture. As more partial ruptures start to appear (Fig. \ref{fig:recurrence-interval-evolution-X2}c), it becomes shorter, and also more chaotic, just as in G. Finally, the evolution of both the number of ruptures and their size distribution resembles closely that of case G (Fig. \ref{fig:recurrence-interval-evolution-X2}b). 

\begin{figure}
    \centering
    \includegraphics[width=\textwidth]{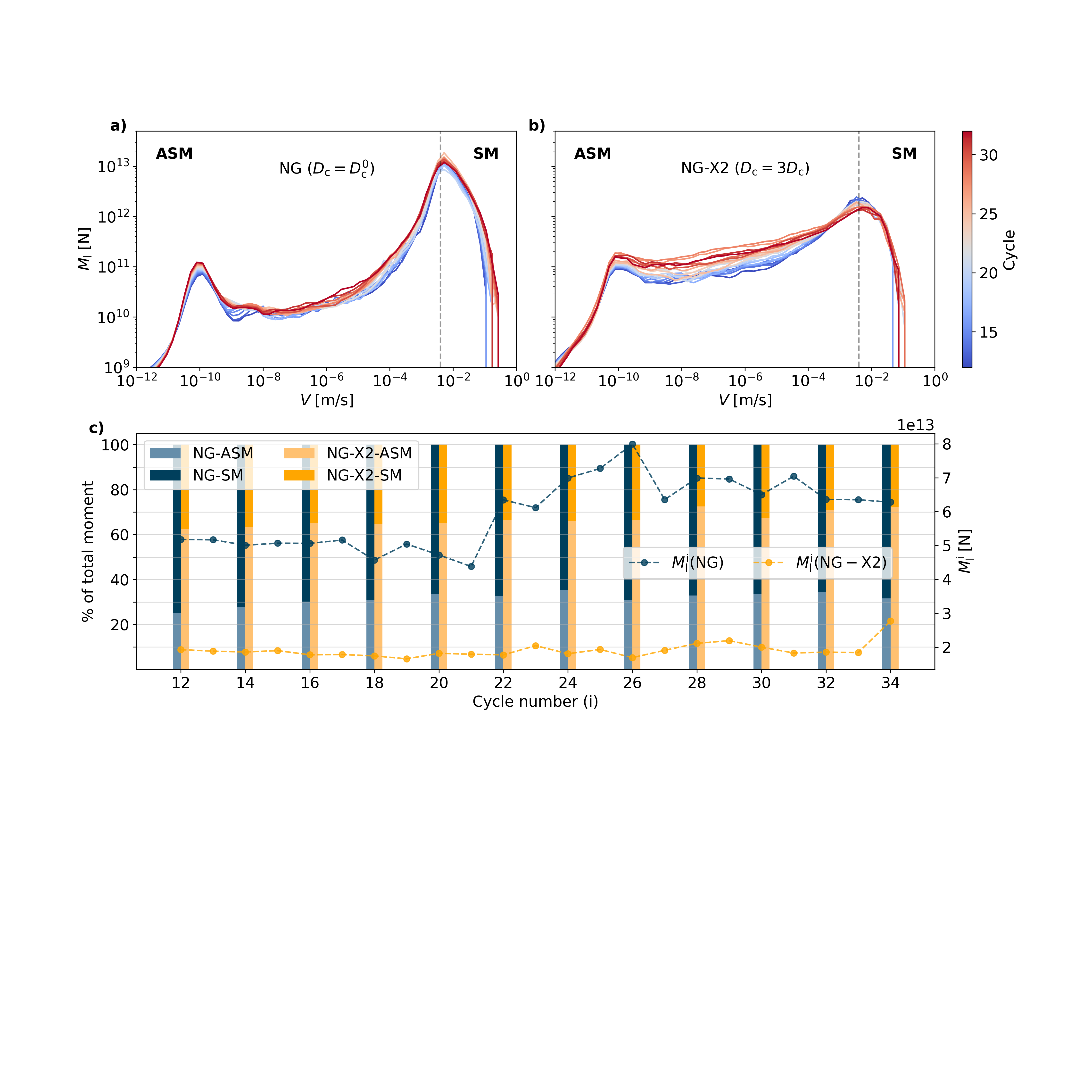}
    \caption{Comparative partitioning of the moment budget per unit length for both cases. \textbf{a, b}) Average distribution of the moment over a whole range of slip rates, aseismic (ASM) and seismic (SM) for cycles 12 to 34 and cases NG (a) and NG-X2 (b). \textbf{c}) Evolution of both slip rate components (Aseismic [ASM] and Seismic [SM]) across cycles (left axis), as well as the moment per cycle (right axis), for both NG and NG-X2 cases. 
    }
    \label{fig:moment-budget-partitioning-X2}
\end{figure}

To conclude, the partitioning of the moment budget is dominated by the aseismic component, which accounts for around 70\% of all released moment, and it remains rather constant throughout the entire sequence, due to the slow evolution of the fault without accumulation of gouge. The total moment per cycle released by NG-X2 is around a third of the one released by NG (see Fig.~\ref{fig:moment-budget-partitioning-X2}c).

\section{Nucleation with higher normal stress (HNS)}

\begin{figure}
    \centering
    \includegraphics[width=\textwidth]{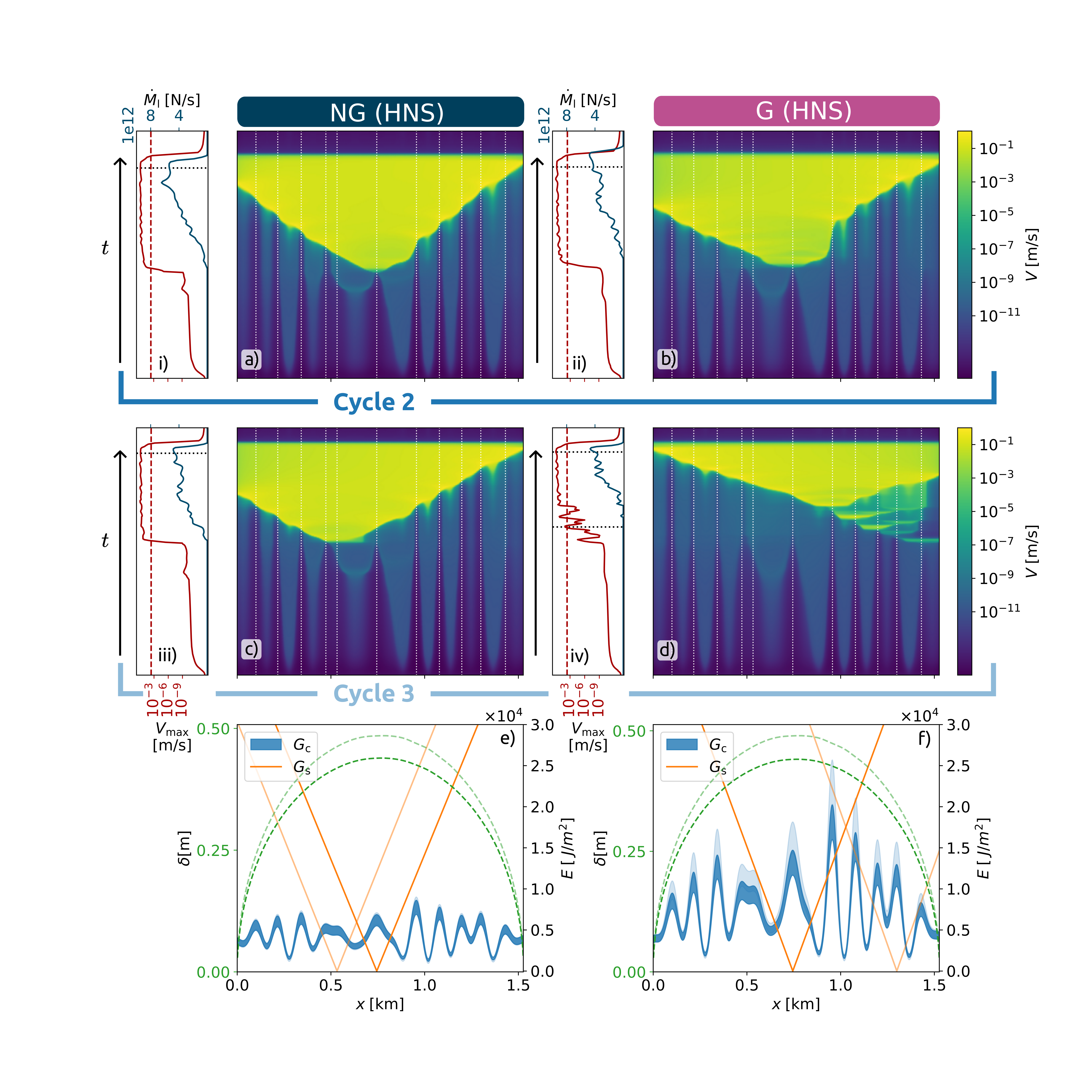}
    \caption{Earthquake nucleation process for representative cycles in NG-HNS and G-HNS. Data for both cases is shown for cycle 2 (a,b) and cycle 3 (c,d) of a given set of simulations.
    For each cycle, the evolution of the maximum fault slip rate (red) and the moment rate per unit length (dark blue) are shown in (i-iv), where earthquakes are signaled with dotted lines. The color maps (a-d) show the evolution of the slip rate distribution over the entire fault from the beginning to the end of the cycle. At the bottom (e,f), approximations for $G_\mathrm{s}(x)$ and $G_\mathrm{c}(x)$ are plotted for both cases, along with the slip distribution $\delta(x)$ (dotted green lines). $G_\mathrm{c}(x)$ is plotted for a range of observed co-seismic slip rates going from $10^{-3}$~m/s to $10^{-4}$~m/s. Data from cycle 2 is shown in dark color, while that from cycle 3 is displayed in lighter color.}
    \label{fig:nucleation-10MPa}
\end{figure}

If we set the average normal stress $\sigma^0_\mathrm{n}$ to 10~MPa, which is a much more realistic value, we are only able to simulate up to 3-4 cycles instead of 30-40, that is, 10 times fewer cycles. This is because the time-step is controlled by $\sigma_\mathrm{min}$, which is artificially set to 1~kPa for all cases, as explained in the main text, so we can only simulate up to 30 years within a reasonable run-time, which in this case (NHS) amounts to around 4 cycles only. However, since the stress is 10 times higher, the amount of gouge produced is comparable (see section 5), but even so, the nucleation dynamics is quite different, with almost no partial ruptures present and little to no precursory aseismic slip (Fig.~\ref{fig:nucleation-10MPa}). This is because, contrary to the NG-X2 case, where the uniform increase of $D_\mathrm{c}$ makes only the fracture energy larger, in this case, the higher normal stress contributes both to the fracture energy and the energy release rate (Eqs.~11 and 12), just like in NG-X1, where the longer wavelength of $G_\mathrm{c}$ ($\lambda_\mathrm{min}$) compensates for its larger amplitude.   
\newpage

\section{Evolution of gouge (High vs. Low normal stress)}

\begin{figure}
    \centering
    \includegraphics[width=\textwidth]{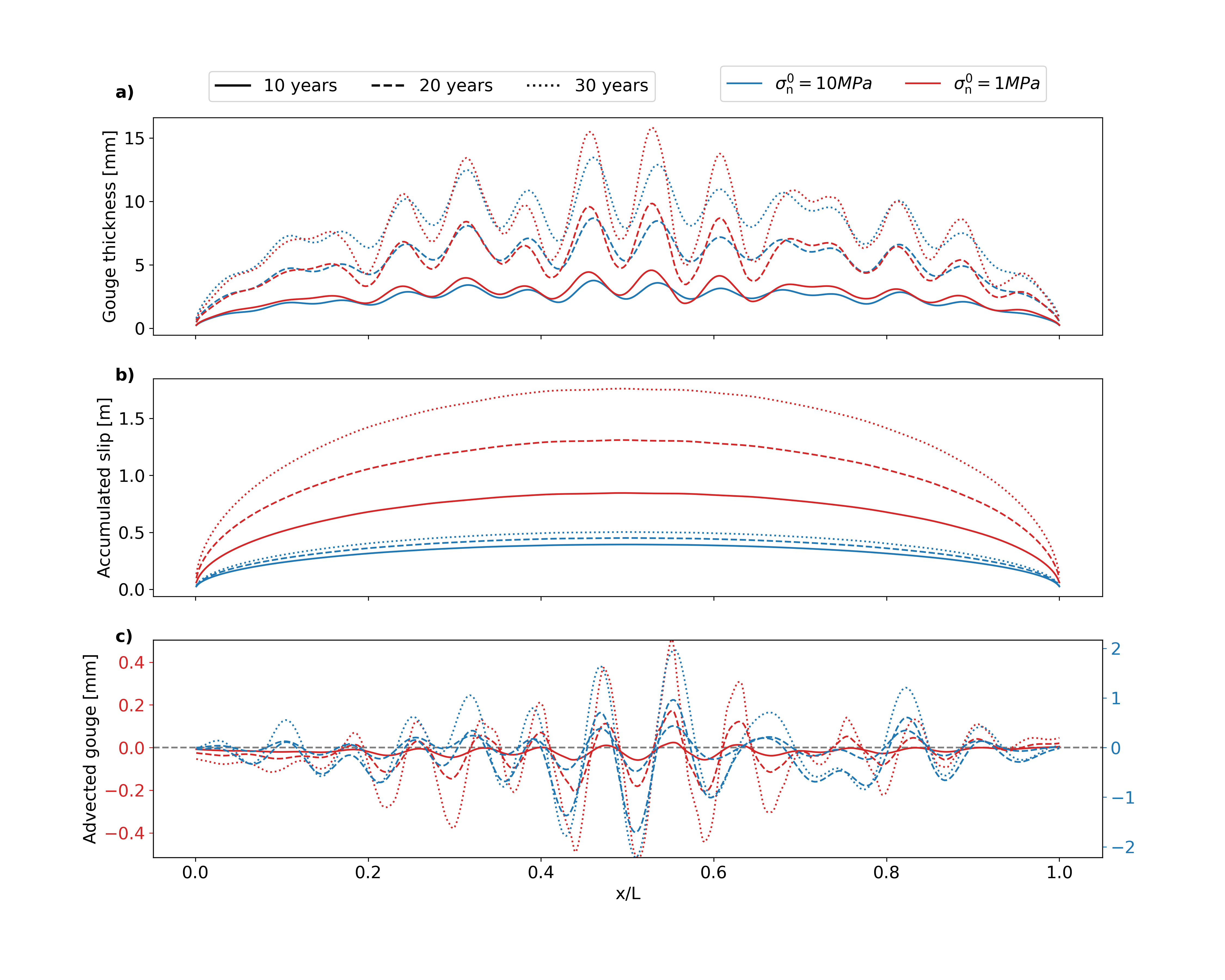}
    \caption{ \textbf{a}) Gouge layer thickness distribution. \textbf{b}) Accumulated slip distribution. \textbf{c}) Advected gouge, either net input (positive) or net output (negative). All data shows the average over the entire set of simulations, 10 (solid line), 20 (dashed line) and 30 years (dotted line) of simulated time and for $\sigma^0_\mathrm{n}=1\mathrm{MPa}$ (blue) and $\sigma^0_\mathrm{n}=10\mathrm{MPa}$ (red). }
    \label{fig:gouge-dist-evolution}
\end{figure}

Lastly, we show the distribution of the gouge layer thickness over time (Fig.~\ref{fig:gouge-dist-evolution}a) for both $\sigma^0_\mathrm{n}=1\mathrm{MPa}$ (blue) and $\sigma^0_\mathrm{n}=10\mathrm{MPa}$ (red). Because the higher normal stress compensates for hosting fewer cycles and therefore, accumulating less slip (see Fig.~\ref{fig:gouge-dist-evolution}b), as mentioned earlier, the amount of gouge produced is similar in both cases, although the amplitude of peaks is more pronounced for $\sigma^0_\mathrm{n}=1\mathrm{MPa}$, reaching an average of up to 15 mm of gouge after 30 years, while only around 13 mm for $\sigma^0_\mathrm{n}=10\mathrm{MPa}$. This is explained by a better re-distribution of gouge through advection in the case with higher normal stress (blue), as can be deduced from Fig.~\ref{fig:gouge-dist-evolution}c, which shows a five-fold difference in advected gouge as compared to the case with lower normal stress. As a result, when the normal stress is higher, more gouge is advected from the peaks of stress, where it is more abundantly produced, to the valleys, where much less is produced, homogenizing its distribution along the fault.    

\newpage

\begin{figure}
    \centering
    \includegraphics[width=\textwidth]{recurrence_interval_evolution_6_X2_v7.png}
    \caption{Evolution of the recurrence interval over multiple cycles. \textbf{a}) Maximum slip rate over time of intermediate cycle (20) for both cases. Vertical dashed lines show the end of the cycle, while horizontal lines indicate the creeping (bottom) and dynamic (top) threshold velocities. \textbf{b}) Evolution of rupture size distribution averaged over simulations in 4-sized groups, together with the $75^{\mathrm{th}}$ percentile. \textbf{c}) Average number of ruptures per cycle. Shaded area shows the standard deviation. \textbf{d}) Evolution of the average recurrence interval across cycles for both NG and NG-X2. Dashed lines show specific cases corresponding to the set of simulations shown in \textbf{a}.}
    \label{fig:recurrence-interval-evolution-X2}
\end{figure}

